\documentclass[comsoc, jounal, 12pt]{IEEEtran}
\usepackage[utf8]{inputenc}
\usepackage[T1]{fontenc}

\usepackage{graphicx}
\usepackage{subfig}
\usepackage{float}

\usepackage[super]{nth}
\usepackage[noadjust]{cite}
\usepackage[shortlabels, inline]{enumitem}

\newcommand{\eg}{\textit{e.g.,~}}
\newcommand{\ie}{\textit{i.e.,~}}

\usepackage{color, xcolor}

\begin{document}

\title{How Far Can We Go in Compute-less Networking: Computation Correctness and Accuracy}

\author{
	\IEEEauthorblockN{
		Boubakr Nour,~\IEEEmembership{Member, IEEE,}
		and Soumaya Cherkaoui,~\IEEEmembership{Senior Member, IEEE}
	}
	
	\thanks{B. Nour and S. Cherkaoui are with INTERLAB Research Laboratory, Faculty of Engineering, Department of Electrical and Computer Science Engineering, Université de Sherbrooke, Sherbrooke (QC) J1K 2R1, Canada (e-mails: boubakr.nour@usherbrooke.ca, soumaya.cherkaoui@usherbrooke.ca).}
}


\maketitle

\begin{abstract}
	Emerging applications such as augmented reality and tactile Internet are compute-intensive and latency-sensitive, which hampers their running in constrained end devices alone or in the distant cloud. The stringent requirements of such application drove to the realization of Edge computing in which computation is offloaded near to users. Compute-less networking is an extension of edge computing that aims at reducing computation and abridging communication by adopting in-network computing and computation reuse. Computation reuse aims to cache the result of computations and use them to perform similar tasks in the future and, therefore, avoid redundant calculations and optimize the use of resources. In this paper, we focus on the correctness of the final output produced by computation reuse. Since the input might not be identical but similar, the reuse of previous computation raises questions about the accuracy of the final results. To this end, we implement a proof of concept to study and gauge the effectiveness and efficiency of computation reuse. We are able to reduce task completion time by up to 80\% while ensuring high correctness. We further discuss open challenges and highlight future research directions.
\end{abstract}


\IEEEpeerreviewmaketitle

\section{Introduction}
\IEEEPARstart{T}{}he current wave of the Internet of Things (IoT) revolution drove massive changes in the nature of applications and users' requirements, which consequently lead to an enormous number of connected devices and huge generated data. Many of these applications are computation-intensive and delay-sensitive by design.
Although cloud computing has shown great resilience in providing an unrestrained amount of computing resources, it is not a suitable model to satisfy stringent low latency requirements, especially when every millisecond counts. Edge computing~\cite{filali2020multi} has been proposed as an extension of cloud computing to overcome this issue. Edge computing aims at moving services from the distant cloud to the near edge servers (\ie closer to end-users) and offloading computational tasks from resource-constrained devices to the edge, and consequently, meet the desired Quality of Experience (QoE). However, it is not feasible to move all cloud's immense resources to the edge for data analysis and treatment.

IoT applications share a many-to-one correlation between the service's input and output data. The same service may have multiple inputs that lead to the same output data~\cite{guo2018foggycache}, even if the inputs are (semi)-similar.
Let us assume a smart tourism example, where a group of tourists visits the Egyptian pyramids. Visitors could benefit from their smartphones to take pictures of various statues and then query related information. Object (statue) recognition is a CPU-intensive task that can ideally be offered by near-edge servers. Yet, the edge will receive different images taken by multiple users through different angles, yielding (after computation) the same output (the same statue).
Another scenario is the use of Virtual Reality (VR) services to experience the environment many centuries ago (\eg LithodomosVR\footnote{LithodomosVR: www.lithodomosvr.com}). Users will use their 3D glasses or headsets to send snapshots of the current view to the near edge, where the VR service is executed. Multiple users located at the same spot can take the same view but from different angles that yield the same computation results.
Tactile Internet goes even beyond the examples mentioned above, by enabling real-time transmission of haptic information. With tactile Internet, it will be possible to experience and interact with remote objects as if they were located locally. Edge computing is a key enabler to meet the ultra-low delays in tactile Internet. Redundant interactions and computations that lead to the same computational results will be extensively used.

These services share many common facts. The first is that all of them require a massive computational capability that cannot be offered by the device itself, high communication bandwidth that cannot be guaranteed by the back-haul network, and ultra-low latency that cannot be provided with the current mobile networks. The second fact is that the edge server computes redundant tasks with identical or (semi)-similar input data that produce the same output \cite{azad2021promise}. The similarity metric refers to the fact that tasks have temporal, spatial, or even semantic correlations between the computation's input and output data. Eliminating these redundant computations by reusing the output results of previous (similar) executed tasks instead of recomputing them from scratch will reduce the execution cost and enable the edge server with low computation capacity to sustain the expected quality of experience.

The computation reuse concept -- can be realized either fully (\ie the previously-stored output is used to satisfy the execution of newly arrived tasks), or partially (\ie the output is used to fully satisfy part of the execution, the rest can be completed via a computation from scratch), is one of the pillar techniques used in compute-less networking~\cite{nour2020computeless}.
A compute-less network is built on top of the edge computing paradigm through a set of communication and computation optimization mechanisms, such as service offloading, in-network computing, task clustering and aggregation, etc. Services are offloaded to the edge server and stored based on their popularity and the edge capacity. Compute-less networking attempts to reduce the amount of computation at the network, minimize the amount of content traversed in the network, meet the strict quality of experience requirements, and lessen resource utilization~\cite{bellal2021coxnet}.

\vspace*{0.2cm}
\textit{Motivation.}
Whatever the advantages of the computation reuse concept in compute-less networking are, producing the same intact output based on different inputs is always a challenge. Even if the surjective property (\ie an output can be obtained via different inputs) of the service could be irrevocable by design, the correctness and accuracy of the final output are dubious especially when relying on partial computation reuse. This research question is the main motivation behind our work, in which we will investigate the correctness and accuracy of the computation reuse concept.

\vspace*{0.2cm}
\textit{Contributions.}
The main contribution of this work is not limited to studying the correctness of computation reuse concept, but also to investigating its effectiveness and efficiency in improving computation and communication.
To the authors' best knowledge, this is the first work that investigates the correctness of computation reuse at the network level. Indeed, this paper has multifold contributions that can be summarized as follows:
\begin{enumerate*}[(i)]
	\item we study the applicability of compute-less networking in different domain-specific applications and real-time systems,
	\item we design a proof of concept to gauge the accuracy and correctness of output data when computation reuse concept is applied, and
	\item we investigate the effectiveness and efficiency of compute-less networking and highlight different issues and future research directions.
\end{enumerate*}

The rest of this paper is organized as follows: in Section II, we introduce the compute-less networking concept. In Section III, we present different scenarios and use-cases that integrate compute-less networking in real-world applications. In Section IV, we elaborate on the computation correctness and accuracy by presenting a proof of concept along with its performance. Section V highlights some issues and provides future research directions, and in Section VI, we conclude our work.

\begin{figure*}[!t]
	\centering
	\includegraphics[width=0.7\linewidth]{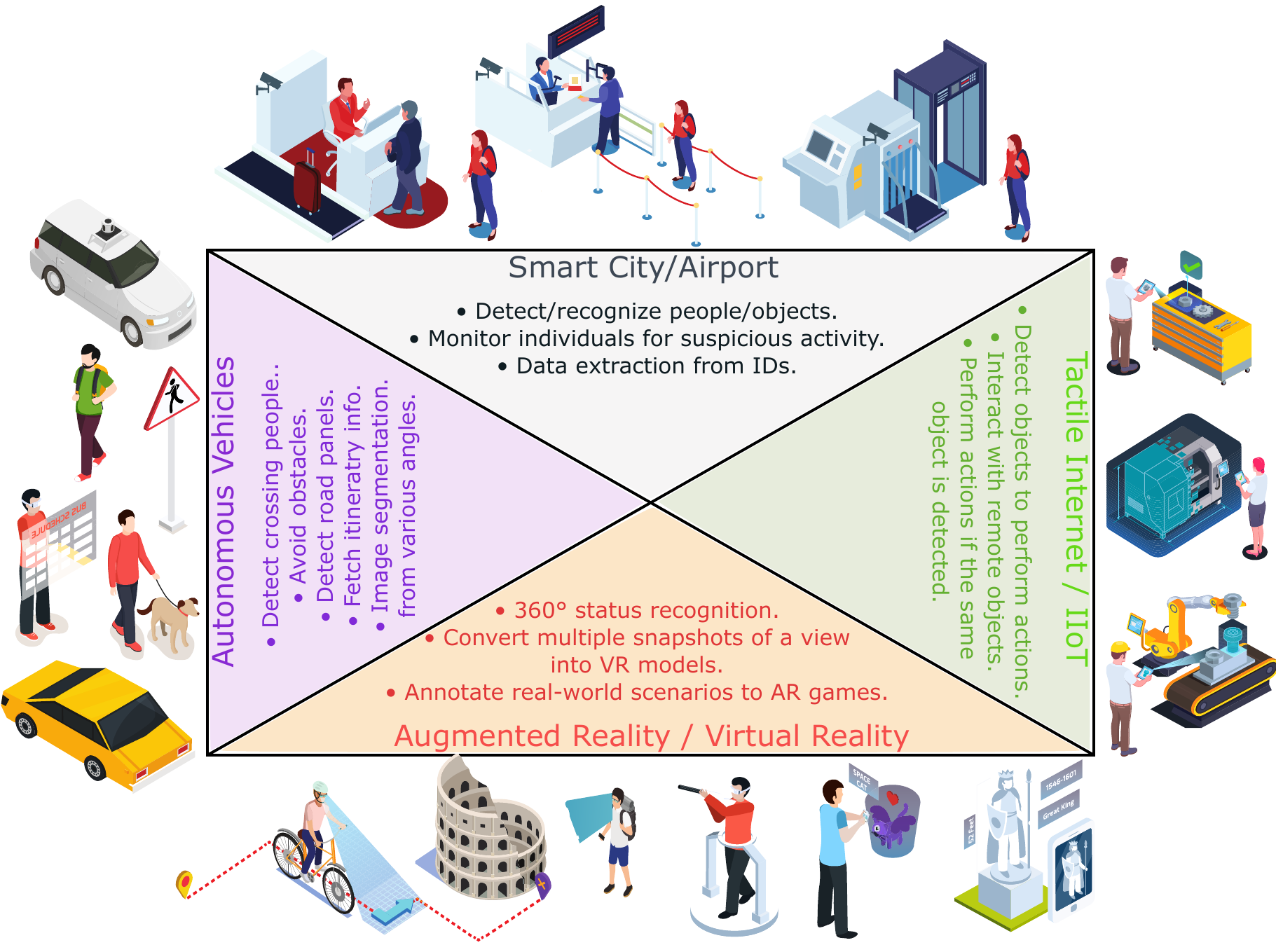}
	\caption{An example of integrated compute-less networking in real-world use-cases.}
	\label{fig:use-cases}
\end{figure*}

\section{Compute-less Networking: A Bird's Eye View}
End-users are now (or will be soon) using various applications that require more computation and short delay (\eg augmented reality, tactile Internet). Yet, their mobile devices do not have enough resources to perform these heavy computations. The edge computing paradigm attempts to meet strict application requirements by moving computation from the distant abundant cloud servers to a near satisfactory edge server~\cite{bellal2021coxnet}.
Other additional paradigms include the integration of networking and computing to further improve response times and the utilization of network and computing resources. Different IETF research groups such as the {\em Computing in the Network Research Group} (COINRG)~\cite{kutscher-coinrg-dir-02} and the {\em Routing Area Working Group} (RTGWG)~\cite{geng-rtgwg-cfn-dyncast-ps-usecase-00} are exploring numerous possibilities to integrate computation with networking. Work in~\cite{krol2019compute} explored the applicability of computing in the network concept and the existing challenges and issues. Researchers in~\cite{li-rtgwg-cfn-framework-00} designed a {\em Compute First Networking} (CFN) architecture that leverages both computing and networking to determine the optimal edge network among multiple edge networks within different geographic locations.

In high-density networks (\eg big cities, Industrial IoT), most services require location-aware computing, where end-users profusely invoke the same service with different input data, and most of these computations have an input-output correlation~\cite{mtibaa2020compute}.
In compute-less networking~\cite{nour2020computeless}, network systems need to perform the absolutely minimum amount of computation and communication. In doing so, we need to ensure that the computation is offloaded to the near edge server and the execution of duplicate computations is highly eliminated. Indeed, compute-less networking is an enhancement of edge and in-network computing~\cite{mai2020network}. This can be enabled through the fact that received tasks for execution may share parts of the required computation in common. The computation reuse concept can be adopted to reuse (partially or fully) the results (final or intermediate) of already executed tasks for the execution of newly received tasks. Computation reuse can contribute to the minimization of the execution of duplicate computation, and hence enhance the quality of service and improve resource utilization. However, identifying if multiple tasks are (semi)-similar is not a straightforward process.
To this end, edge servers need to identify tasks, store the results of previously executed tasks, and efficiently search for parts of these tasks that can be reused for the execution of each newly received task~\cite{nour2021whispering}.

\section{Scenarios and use-cases}
As depicted in Figure~\ref{fig:use-cases}, compute-less networking can be integrated into different real-world use-cases, including but not limited to smart cities, autonomous vehicles, augmented reality, and tactile Internet. In the following, we present a review of different use-cases that can benefit from compute-less networking paradigm.

\vspace*{0.2cm}
\textit{Smart Cities.}
Smart and autonomous services are widely expanded in the realization of smart cities. Smart devices such as cameras are essentially deployed for people/objects detection and recognition. In an airport, for example, cameras are used in terminals and gates to monitor individuals for unusual activity in the boarding process and identify suspicious passengers. The automated ID control process is also used to automatically detect the information from ID using {\em Optical Character Recognition} (OCR). The OCR system can further be employed to detect handwriting and then provide panels and text translation to different languages. The common thing among these use-cases is that the same process can be done multiple times yielding the same output~\cite{nour2020computeless}.

\vspace*{0.2cm}
\textit{Autonomous Vehicles.}
Let us consider a scenario where self-driving vehicles use cameras to detect crossing people and avoid obstacles around so that they navigate without human intervention towards the destination. In doing so, the vehicle relies on object detection services offered by the near edge server. Multiple cameras and sensors deployed on different vehicles may take the same snapshot for the same object(s) but from different angles. Road panels detection and recognition follow the same concept, where all vehicles on the same driving lane/road are required to detect the panel's information. The same remark can be applied to image segmentation, in which the service aims to segment vehicles, bicycles, pedestrians, obstacles, sidewalks from the same snapshot.  All these services are executed multiple times but using similar data~\cite{mtibaa2020compute}.

\vspace*{0.2cm}
\textit{Augmented Reality.}
Augmented Reality (AR) allows sweeping our daily lives via a variety of mobile-based services. For instance, services such as Google Daydream, Earth VR, and Lithodomos VR enable virtual content like 3D models, animations, and annotations to be placed on top of a real-world environment. Similarly, mobile-based AR games, such as Pokémon GO, allow players to use their location to challenge other players by enhancing the game with real and virtual world interactions using the smartphone camera. These services are usually invoked via similar inputs (\eg view in ancient cities, spots in big cities, etc)~\cite{shannigrahi2020next}.

Regardless of the context of the provided service, the common factor between these use-cases is that the service will most likely receive multiple invocations with different input data that produce the same output. The inputs of these invocations are similar, either partially or fully, which implies that a common computation is executed redundantly. Yet, using partial computation may mislead the intact computation output. In the next section, we present a proof of concept to study the correctness and accuracy of computation reuse.

\section{Computation Correctness: A Proof of Concept}
In this section, we study the correctness when the computation reuse concept is adopted. We start by describing the experiment environment and then discussing the obtained results.

\begin{figure}[!t]
	\centering
	\includegraphics[width=\linewidth]{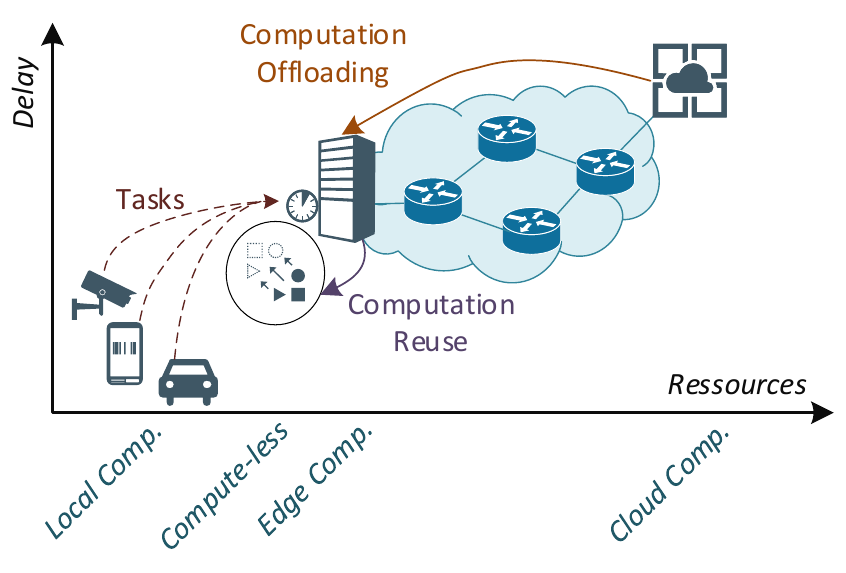}
	\caption{Evaluation network.}
	\label{fig:evaluation_network}
\end{figure}

\vspace*{0.2cm}
\textbf{Experimental Setup.}
The main objective of this work is to study if the computation reuse concept will lead to an accurate computation output similar to what computation from scratch achieves. 
In doing so, we implemented the Proof of Concept shown in Figure~\ref{fig:evaluation_network}.
We consider an \textit{object detection use-case} that can be either executed at the edge or cloud server. A list of end-users sends up to 100 tasks, following the Zipf distribution. Tasks carry different images, which are considered input data for object detection. The service is offloaded to the edge network to meet the computation delay constraints. The cloud server is located 8 hops from end-users with significant resources, while the edge server is located only 2 hops away with fewer resources compared with the cloud server. The further away the tasks are forwarded, the more resources are available, but the largest delay occurs.
Additionally, we extended the edge server with a compute-less paradigm by performing computation reuse. The edge server keeps track of previously executed tasks in a Reuse Table by storing the task's input and output data, which is also used to complete the execution of incoming tasks. To keep the Reuse Table compact, we applied {\em Least Frequency Used} (LFU) policy to clean the table by removing the entries that have a smaller frequency usage rate. The computation can be done in three different models:

\begin{itemize}
	\item \textit{At the cloud}: all tasks are sent to the cloud server for remote execution. Although the cloud has limitless computing capacity, it is located far away from end users.
	
	\item \textit{At the edge}: all tasks are executed at the edge server that is located near to the end-user, yet has fewer computing resources compared with the cloud server.
	
	\item \textit{At the edge with computation reuse}: the edge performs computation reuse for each newly received task, if applicable. Otherwise, computation from scratch is performed.
\end{itemize}

\begin{figure}[!t]
	\centering
	\includegraphics[width=\linewidth]{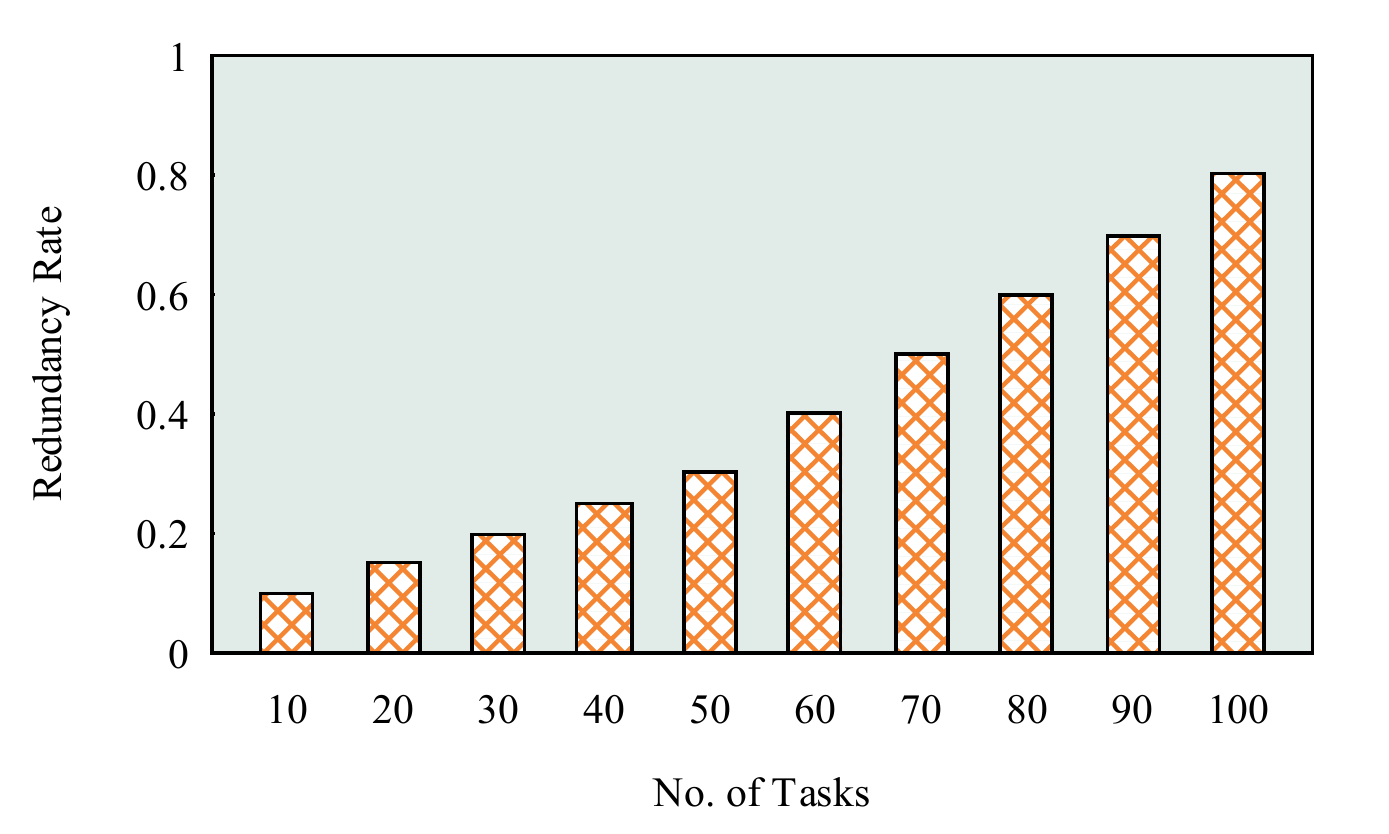}
	\caption{Redundancy Rate.}
	\vspace{-0.5cm}
	\label{fig:redundancy_rate}
\end{figure}

In compute-less networking, when an edge server receives a new task, the server checks first if a (semi)-similar match exists on its local Reuse Table. We implemented {\em Locality-Sensitive Hashing} (LSH) algorithm that tends to hash similar input data to the same bucket.
The space complexity of LSH is $O(lnk)$, where $l$ is the number of hash tables, $n$ number of elements in each table, and $k$ is the dimension of hash vector. By ignoring lower-order terms, the space complexity is $O(n^{1+\rho})$, where $\rho$ determines the time/space bounds of LSH algorithm. The time complexity to find a match in a hash table is $O(lk)$. For each candidate, we spend $O(d)$ time to compute the distance. Ignoring lower-order terms, the query time is $O(n^{\rho})$.
If a match is found, the stored output will be used as an output of the received task without performing any extra computation. Otherwise, a computation from scratch will be performed. 
Probabilistic models and machine learning techniques can be applied to improve the matching ratio. It is important to highlight that the same task can use multiple computation reuse outputs from different previously-stored tasks (\eg segmenting an image into multiple parts and each segment has an object that is stored on the Reuse Table).

We used the standard image dataset, \textit{ImageNet}~\cite{ILSVRC15}. The dataset contains over 4000 types of labeled images taken from different viewpoints and lighting conditions. We used the {\em Yolo} framework for object detection service~\cite{redmon2016you}. End-users send selected images from the dataset with a redundancy rate. Figure~\ref{fig:redundancy_rate} depicts the redundancy rate for each set of requested tasks. This metric refers to how many times an object frequently appears in the list of requested images received by the server, which can be either a redundant image in the list (\ie full redundancy) or the same object in multiple images (\ie partial redundancy). In the following, we present the \nth{90} percentile of the results collected after 10 trials.

\vspace*{0.2cm}
\textbf{Results and Discussions.}
To evaluate the correctness of computation reuse and to gauge the efficiency of compute-less networking, we consider four main metrics: task completion time, task computation time, correctness rate, and completion gain.

\begin{figure}[!t]
	\centering
	\includegraphics[width=\linewidth]{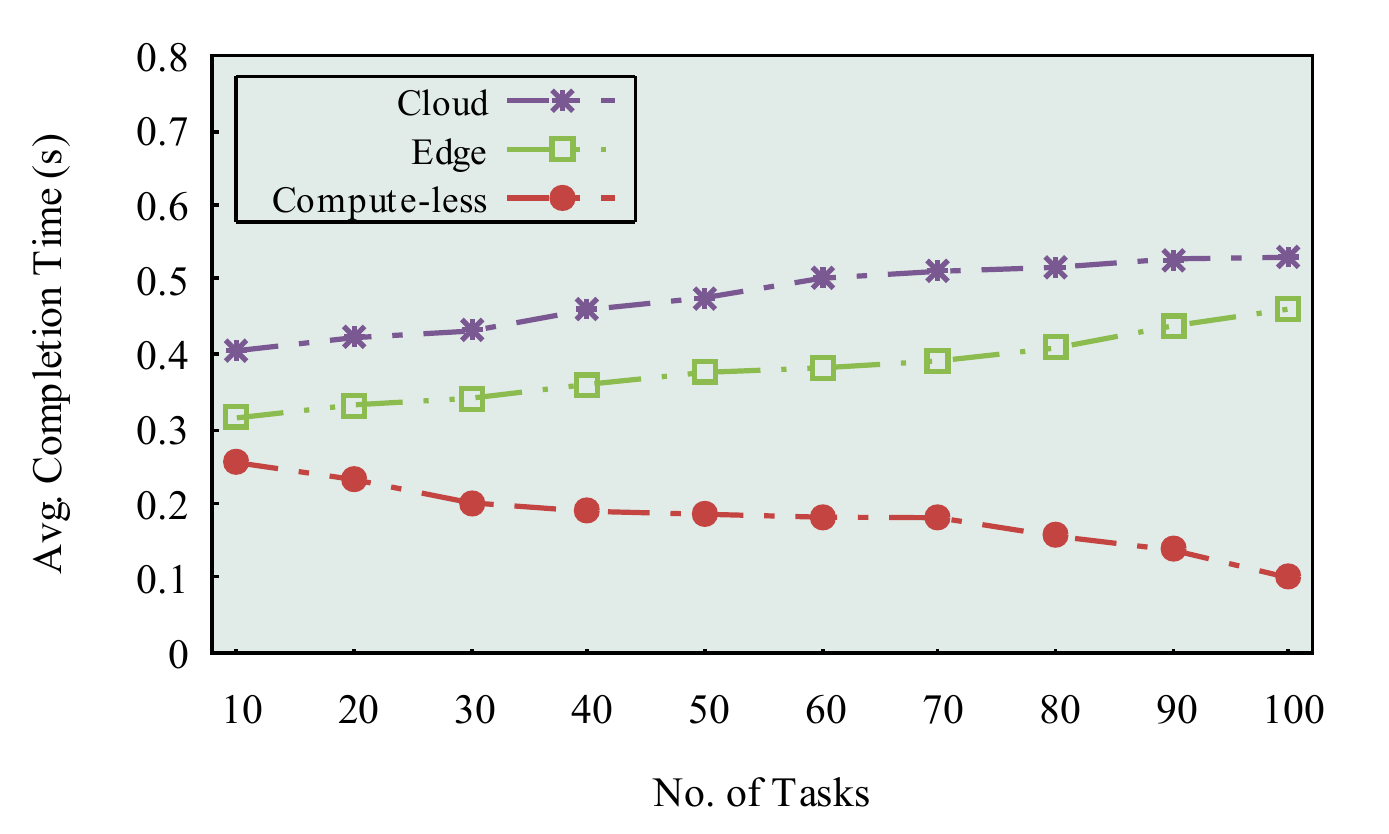}
	\caption{Task completion time.}
	\vspace{-0.5cm}
	\label{fig:completion_time}
\end{figure}

\vspace*{0.1cm}
\textit{Task completion time.} It is measured as the overall time elapsed between end-users issuing their tasks for execution and receiving back the execution results.
Figure~\ref{fig:completion_time} depicts the average task completion time. Cloud has the largest completion time since the server is located far away from end-users, and transmitting a large size of input data takes more time. On the other hand, as the edge server is located near to end-users (\eg one hop), we notice that the completion time is reduced compared to the cloud since we eliminated the transmission of a huge amount of data in the network, yet the edge server has a significant waiting time to execute all received tasks. However, by adopting computation reuse, we can observe that the completion time is much better than for edge computing, and decreased when the number of tasks increased (up to $80\%$). This is due to the fact that the received tasks witness a growing redundancy rate that accelerates the completion time. When the redundancy rate (Figure~\ref{fig:redundancy_rate}) reaches 0.8, the completion time for compute-less networks decreases up to $0.08~(s)$ compared to $0.27~(s)$ for edge computing.

\begin{figure}[!t]
	\centering
	\vspace{-0.6cm}
	\includegraphics[width=\linewidth]{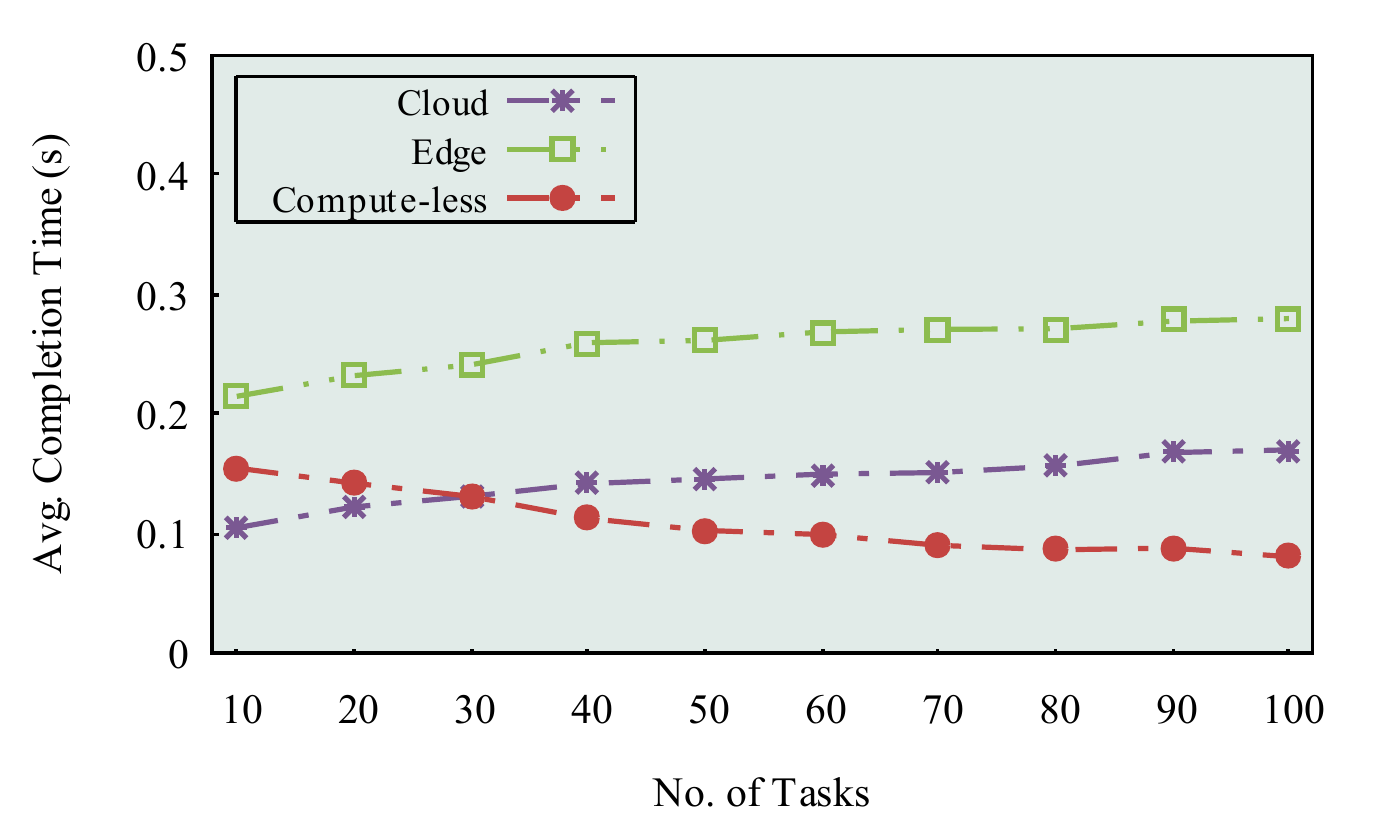}
	\caption{Computation time.}
	\vspace{-0.5cm}
	\label{fig:computation_time}
\end{figure}

\vspace*{0.1cm}
\textit{Task computation time.} It is measured as the overall time elapsed between receiving the task's input data (at the server) and producing the final output. 
Figure~\ref{fig:computation_time} outlines the average computation time. Indeed, this result proves that the transmission time has a high impact on the performance of cloud computing even with unlimited computation resources, while the waiting time impacts the edge server capacity to perform fast computation. Yet, the waiting time at the edge is negligible compared to the transfer time for the cloud (refer to Figure~\ref{fig:completion_time}), which proves the outperformance of the edge compared to the cloud. On the other hand, compute-less exceeds even the cloud performance by processing a very negligible lookup process to find a match on the Reuse Table. This is due to the fact that services have a high complexity that requires resources (at the cloud), while the lookup process is weightless (with compute-less). When the redundancy of received inputs is high, compute-less witnesses a very fast computation since it finds a quick match compared to computing the task even with the immense resources at the cloud server.

\begin{figure}[!t]
	\centering
	\includegraphics[width=\linewidth]{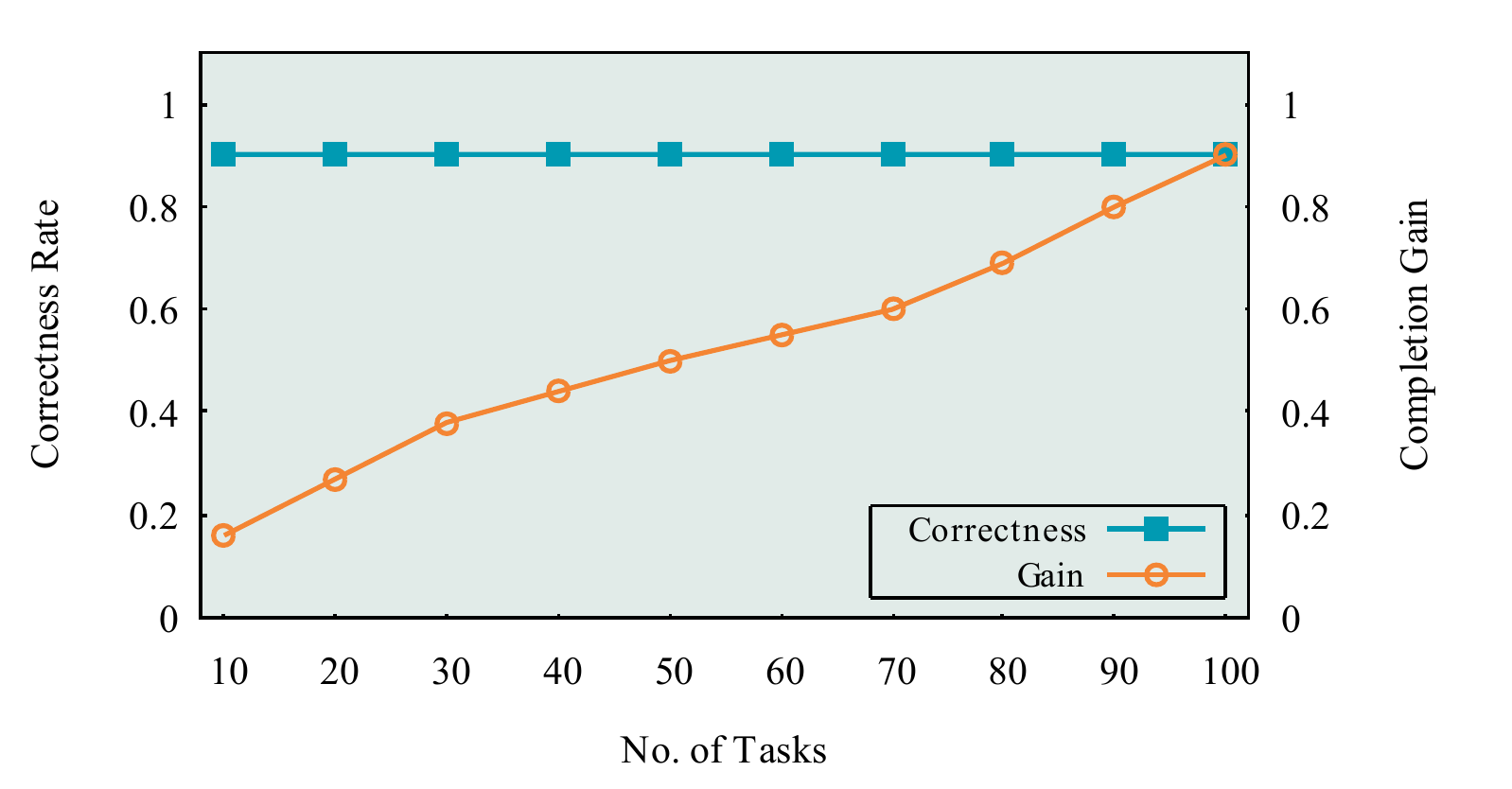}
	\caption{Computation correctness and gain.}
	\vspace{-0.5cm}
	\label{fig:correctness_rate}
\end{figure}

\vspace*{0.1cm}
\textit{Correctness Rate and Completion Gain.} 
The correctness refers to the percentage of achieving accurate output data (results) similar to computation from scratch. The highest the correctness, the better the accuracy. The completion gain refers to the non-utilized resources by adopting computation reuse compared with computation from scratch at the edge.
Figure~\ref{fig:correctness_rate} shows that compute-less networking is able to reach almost 0.9 of correctness rate, which means 90\% of the reuse operations are absolutely correct. The missed 0.1 reverts to the fact that some of the computations have not been stored in the Reuse Table or have been evicted by the replacement policy. Hence, there is a need to compute them from scratch. This means that the final output is reached via a mix of computation reuse and a computation from scratch.  
On the other hand, the gain is increased when the number of tasks increases. The redundancy factor of received inputs has a direct impact on this result since high redundancy implies more computation reuse, and consequently more gain and less resource utilization.

\section{Issues and Future Directions}

\vspace*{0.05cm}
\textbf{Input Similarity Detection.}
The functions' input values are rarely identical. Yet, they can be correlated temporally, spatially, or semantically, and mapped to the same output. Finding if multiple inputs are similar drives to find the most similar record (\ie distance measure) that is the nearest to an input, known as the nearest neighbor. This process is often computationally costly and time-consuming. Therefore, numerous methods have been proposed to undertake this issue. Hashing is one of the common solutions. Locality-sensitive hashing, indeed, aims at regrouping similar inputs into the same buckets based on hash functions that map a similar input to the same hash code with a higher probability of dissimilar inputs. Near neighbor search techniques can also be used to find the nearest similar input in a set of data.

\vspace*{0.2cm}
\textbf{Multiple Inputs Aggregation.}
Input data aggregation process is an essential constituent for data management in modern services, which is core to enhancing the quality of deployed service and decreasing the design effort. Inputs data aggregation is a built-in pre-process on (some or all) functions that form the service. It is defined as the process of producing synthesized forms from multiple input data items. Its main goal is to provide a compact data format easily processed by the function. In-network input data aggregation will help to reduce the amount of transmitted data (less communication) while ensuring better resource exploitation.

\vspace*{0.2cm}
\textbf{Microservice-based Applications:}
Serverless technology has recently started gaining more attention due to the enormous features it affords such as scalability, reliability, and pay-as-you-go with affordable computational. A serverless-based application is no more than a distribution of loosely-coupled functions among different edge servers. These functions work in coordination with each other to achieve the desired tasks aiming at reducing the development cost and deployment effort. These types of applications usually share some functions with each other. The duplication of these functions in various locations (edge servers) is also possible to withstand failures and ensure high availability. This will contribute perfectly to computation reuse opportunities since shared functions are likely more requested compared to others, which increases the reuse gain. Moreover, the computation reuse of a serverless application will not only depend on its previous computations but also on the other applications' computations since they are all sharing the same functions to build the complete service.

\vspace*{0.2cm}
\textbf{Input/Data Privacy.}
As far as data privacy is concerned, end-users should authorize the service provider to store their input/output data at the edge server and be able to reuse it to satisfy other computations. The same edge server can accommodate multiple services offered by multiple providers (\eg shared edge server), computation sharing across these services is also feasible and will help in improving the quality of experience. Similarly, computation sharing among multiple nearby-edge servers is worthwhile instead of forwarding the request to a distant cloud. In either example, it is mandatory to deploy an efficient data privacy scheme that enables the server to take advantage of the existing or nearby stored information without concern for user/data privacy.

\section{Conclusion}
\label{sec:conclusion}
Cloud computing has given great support to CPU-intensive applications by providing immense computing resources. Emerging applications require strict constraints such as a low response delay, for which the cloud model is no longer suitable. Offloading the computation toward the edge network, closer to consumers, is promising to address this issue. Yet, the network still witnesses heavy amounts of computation and communication closer to the edge. Compute-less networking promises to reduce the amount of computation and communication by eliminating redundant computational {at the network level} through the computation reuse concept. This paper is the first study of its kind, which emphasizes the correctness and accuracy of computation reuse and the efficiency of compute-less networking. To this end, we designed a proof of concept where we were able to reduce the task completion time by up to 80\% while ensuring high correctness.

\section{Acknowledgments}
The authors would like to thank the Natural Sciences and Engineering Research Council of Canada (NSERC) for the financial support of this research.

\bibliographystyle{IEEEtran}

\section*{Biographies}

\vskip -2.0\baselineskip plus -1fil

\begin{IEEEbiographynophoto}{Boubakr Nour}
	is currently working as a Post-Doctoral Research Fellow in INTERLAB Research Laboratory, at the Department of Electrical and Computer Engineering, Université de Sherbrooke, Canada. He received his Ph.D. degree in Computer Science and Technology at the Beijing Institute of Technology, Beijing, China. His research interests include next-generation networking and Internet. He is the recipient of the Best Paper Award at IEEE GLOBECOM (2018), and the Excellent Student Award at Beijing Institute of Technology in 2016, 2017, and 2018 consecutively.
\end{IEEEbiographynophoto}

\vskip -2.0\baselineskip plus -1fil

\begin{IEEEbiographynophoto}{Soumaya Cherkaoui}
	is a Full Professor at Department of Electrical and Computer Engineering, Université de Sherbrooke, Canada. Her research and teaching interests are in wireless networks. Particularly, she works on next generation networks Edge computing/Network Intelligence, and communication networks. Since 2005, she has been the Director of INTERLAB, a research group which conducts research funded both by government and industry. Before joining U. Sherbrooke, she worked for industry as a project leader on projects targeted at the Aerospace Industry. Her work resulted in technology transfer to companies and to patented technology. She has published over 200 research papers in reputed journals and conferences. She has been on the editorial board of several journals including IEEE JSAC, IEEE Network, and IEEE Systems. Her work was awarded with recognitions including a best paper award at IEEE ICC in 2017. She is currently an IEEE ComSoc Distinguished Lecturer. She is a Professional Engineer in Canada and has been designated Chair of the IEEE ComSoc IoT-Ad hoc and Sensor Networks Technical Committee in 2020.
\end{IEEEbiographynophoto}
\end{document}